\newread\testifexists
\def\GetIfExists #1 {\immediate\openin\testifexists=#1
	\ifeof\testifexists\immediate\closein\testifexists\else
	\immediate\closein\testifexists\input #1\fi}
\def\epsffile#1{Figure: #1} 	

\immediate\openin\testifexists=e
	\ifeof\testifexists\immediate\closein\testifexists\else
	\immediate\closein\testifexists\input e\fipsf 

\magnification= \magstep1	
\tolerance=1600 
\parskip=5pt 
\baselineskip= 5 true mm \mathsurround=1pt
\font\smallrm=cmr8

\font\medrm=cmr9

\font\bigbf=cmbx12
 	\def\Bbb#1{\setbox0=\hbox{$\tt #1$}  \copy0\kern-\wd0\kern .1em\copy0} 
	\immediate\openin\testifexists=a
	\ifeof\testifexists\immediate\closein\testifexists\else
	\immediate\closein\testifexists\input a\fimssym.def 
\def\secbreak{\vskip12pt plus .5in \penalty-200\vskip 0pt plus -.4in} 
   \def\newsect#1{\secbreak\noindent{\bf #1}\medskip}
\def\hugeskip{\vskip9mm plus 3mm}
\def\Narrower{\par\narrower\noindent}	
\def\Endnarrower{\par\leftskip=0pt \rightskip=0pt} 
	\def\ra{\rightarrow}		
\def\a{\alpha}  	\def\b{\beta}  	\def\g{\gamma}  
\def\d{\delta}  	 	\def\e{\varepsilon}

     	\def\j{\psi}   	
\def\r{\varrho} 	\def\s{\sigma}

\def\HH{{\cal H}}

\def\cl{\centerline}	
\def\ni{\noindent}  	\def\pa{\partial} 	\def\dd{{\rm d}}
\def\tl{\tilde}     	\def\bra{\langle} 	\def\ket{\rangle}
 
\def\fn#1{\ifcase\noteno\def\fnchr{*}\or\def\fnchr{\dagger}\or\def
	\fnchr{\ddagger}\or\def\fnchr{\medrm\S}\or\def\fnchr{\|}\or\def
	\fnchr{\medrm\P}\fi\footnote{$^{\fnchr}$} 
	{\scrunch#1\toe}\ifnum\noteno>4\global\advance\noteno by-6\fi
	\global\advance\noteno by 1}
	\def\scrunch{\baselineskip=10 pt \medrm}
	\def\toe{\vphantom{$p_\big($}}
	\newcount\noteno

\def\ffract#1#2{{\textstyle{#1\over#2}}}
\def\fract#1#2{\raise .35 em\hbox{$\scriptstyle#1$}\kern-.25em/
	\kern-.2em\lower .22 em \hbox{$\scriptstyle#2$}}

\def\half{\ffract12} 

\def\part#1#2{{\partial#1\over\partial#2}} 
 \def\ref#1{${\vphantom{)}}^#1$}

\def\bbf#1{\setbox0=\hbox{$#1$} \kern-.025em\copy0\kern-\wd0
	\kern.05em\copy0\kern-\wd0 \kern-.025em\raise.0433em\box0}

\def\ref#1{${\,}^{\hbox{\smallrm #1}}$}

\def\Gbar{\raise.13em\hbox{--}\kern-.35em G}
\def\lap{\setbox0=\hbox{$<$}\,\raise .25em\copy0\kern-\wd0\lower.25em\hbox{$\sim$}\,}
\def\glt{\setbox0=\hbox{$>$}\,\raise .25em\copy0\kern-\wd0\lower.25em\hbox{$<$}\,}
\def\gap{\setbox0=\hbox{$>$}\,\raise .25em\copy0\kern-\wd0\lower.25em\hbox{$\sim$}\,}

{\ }\vglue 1truecm
\rightline{SPIN-1999/07}
\rightline{gr-qc/9903084}
\hugeskip
\cl{\bigbf QUANTUM GRAVITY AS A}\medskip
\cl{\bigbf DISSIPATIVE DETERMINISTIC SYSTEM}

\hugeskip

\cl{Gerard 't Hooft }
\bigskip
\cl{Institute for Theoretical Physics}
\cl{University of Utrecht, Princetonplein 5}
\cl{3584 CC Utrecht, the Netherlands}
\smallskip
\cl{and}
\smallskip
\cl{Spinoza Institute}
\cl{Postbox 80.195}
\cl{3508 TD Utrecht, the Netherlands}
\smallskip\cl{e-mail: \tt g.thooft@phys.uu.nl}
\cl{internet: \tt http://www.phys.uu.nl/\~{}thooft/	}
\bigskip
\ni{\bf Abstract}\Narrower

It is argued that the so-called holographic principle will obstruct attempts to produce physically
realistic models for the unification of general relativity with quantum mechanics, unless determinism in
the latter is restored. The notion of time in GR is so different from the usual one in elementary
particle physics that we believe that certain versions of hidden variable theories can \hbox{-- and must
--} be revived. A completely natural procedure is proposed, in which the dissipation of information plays
an essential role. Unlike earlier attempts, it allows us to use strictly continuous and differentiable
classical field theories as a starting point (although discrete variables, leading to fermionic degrees
of freedom, are also welcome), and we show how an effective Hilbert space of quantum states naturally
emerges when one attempts to describe the solutions statistically. Our theory removes some of the
mysteries of the holographic principle; apparently non-local features are to be expected when the quantum
degrees of freedom of the world are projected onto a lower-dimensional black hole horizon. Various
examples and models illustrate the points we wish to make, notably a model showing that massless, non
interacting neutrinos are deterministic.

\Endnarrower \hugeskip

\newsect{1. Introduction.} At present, many elementary particle physicists appear to agree that
superstring theory\ref1 and its descendants such as ``M-theory"\ref2 are the only candidates for a
completely unified theory that incorporates the gravitational force into elementary particle physics.
This concensus is based on the very rich mathematical structure of these theories that shows some
resemblance to the observed mathematical structure of the Standard Model as well as that of General
Relativity.

It also appears to be a satisfactory feature\ref3 of these theories, that they manage to reproduce the
so-called `holographic principle'\ref4. This principle states that any complete theory combining quantum
mechanics with gravity should exhibit an upper limit to the total number of independent quantum states
that is quite different from what might be expected in a quantum field theory: it should increase
exponentially with the surface area of a system, rather than its volume.

Yet this only adds to the suspicion that these theories are far removed from a description of what one
might call `reality'. One would have expected that the quantum degrees of freedom can be localised, as in
a quantum field theory, but this cannot really be the case if theories with different dimensionalities
are being mapped one onto the other. How can notions such as causality, unitarity, and local Lorentz
invariance make sense if there is no trace of `locality' left? In this paper, a theory is developed that
will {\it not\/} postulate the quantum states as being its central starting point, but rather classical,
deterministic degrees of freedom. Quantum states, being mere mathematical devices enabling physicists to
make statistical predictions, will turn out to be derived concepts, with a not strictly locally
formulated definition.

At the time this is written, the quantum mechanical doctrine, according to which all physical states form
a Hilbert space and are controlled by non-commuting operators, is fully taken for granted in string
theory. No return to a more deterministic description of ``reality" is being foreseen; to the contrary,
string theorists often give air to their suspicion that the real world is even crazier than quantum
mechanics. Consequently,  the description of what really constitutes concepts such as space, time,
matter, causality, and the like, is becoming increasingly and uncomfortably obscure. By many, this is
regarded as an inescapable course of events, with which we shall have to learn to live.

But there are also other difficulties associated to such starting points, for instance when space-time
curvature is being used to close an entire universe. We get ``quantum cosmology". An extremely important
example of a quantum cosmological model, is a model of gravitating particles in 1 time, 2 space
dimensions\ref5. Here, a complete formalism for the quantum version at first sight seems to be
straightforward\ref6, but when it comes to specifying exact details, one discovers that we cannot
rigorously define what quantum mechanical amplitudes are, what it means when it is claimed that ``the
universe will collapse with such-and-such probability", what and where the observers are, what they are
made of, and so on. Yet such questions are of extreme importance if one wants to check a theory for its
self-consistency, by studying unitarity, causality, etc.

Since the entire hamiltonian of the universe is exactly conserved, the ``wave function of the universe"
is in an exact eigenstate of the hamiltonian, and therefore, the usual Schr\"odinger equation is less
appropriate than the description of the evolution in the so-called Heisenberg representation. Quantum
states are space-time independent, but operators may depend on space-time points -- although only if the
location of these space-time points can be defined in a coordinate-free manner!\fn{Note that, besides
energy, also total momentum and angular momentum of the universe must be conserved (and they too must be
zero).}

We have learned to live with the curious phenomenon that our wave functions can be eigenstates of
operators which at different space-time points usually do not commute. A ``physical state" can be an
eigenstate of an arbitrary set of mutually commuting operators, but then other operators are not
diagonalized, and so, these observables tend to be smeared, becoming ``uncertain". The idea that such
uncertainties may be due to nothing other than our limited understanding of what really is going on, has
become unpopular, for very good reasons. Attempts at lifting these uncertainties by constructing theories
with `hidden variables', have failed. It is the author's suspicion, however, that these hidden variable
theories failed because they were based far too much upon notions from everyday life and `ordinary'
physics, and in particular because general relativistic effects have not been taken into account
properly. The interpretation adhered to by most investigators at present is still not quite correct, and
a correct interpretation is crucial for making further progress at very technical levels in quantum
gravity.

Earlier attempts by this author to obtain further insights led to the idea that space, time, and matter
all had to be discrete\ref7. If this were the case, it would seem to be easy to set up a deterministic
model of the universe, and a mathematically rigorous procedure to handle probabilities by introducing an
{\it auxiliary\/} Hilbert space, spanned by all possible states, whose evolution is accurately described
by an evolution operator, leading to Schr\"odinger's equation in the continuum limit. Indeed, some models
constructed along these lines look very much like genuine quantum field theories.

They showed, however, one very important shortcoming. This is the fact the the hamiltonian, emerging
naturally from the basic equations, invariably fails to have a lower bound, and so it appeared to be
impossible to construct the vacuum state. One possible exception is a model of (second quantized)
non-interacting massless fermions. They can be viewed {\it exactly\/} as a continuum limit of a discrete,
deterministic theory, see Appendix A. Here it is shown that massless non-interacting neutrinos are
deterministic. Unfortunately, however, we have been unable to generalize this system into something more
interesting.

Since quantum mechanics is described by a {\it unitary\/} evolution operator, it was natural to
expect that, in a cellular automaton model, only time-reversible evolution laws would be
acceptable. However, a little bit of thought suffices to realise that this is not the case. If an
evolution law is not time reversible, it just means that some states will be absolutely forbidden
(their amplitudes will vanish), and others will evolve into states that, after a while, become
indistinguishable from states with a different past. If a pair of states evolve in such a way
that their futures are identical, then these states should be called physically identical from
the very start. To be precise, we must introduce {\it equivalence classes\/} of states, defined
by collecting all states which some time in the future, after a given lapse of time, will become
identical to one another. The evolution from one equivalence class to a different equivalence
class is then again unitary, by construction.

An early attempt to construct a deterministic model with built-in information loss, appeared to bring
improvement: in a certain approximation, the hamiltonian did appear to develop a lower bound\ref8.
Nevertheless, there were shortcomings, as in more precise calculations the lower bound disappears again,
and anyway, the model was unattractive. On the other hand, once it is realized that, {\it at the
classical level}, information loss is permitted, we can return to strictly continuous underlying
deterministic equations. All that is needed is that the equivalence classes are discrete. At later stages
of the theory, one might reconsider the option to regard the continuous theories as the continuum limit
of some discrete system.

The advantages of returning to continuum theories are numerous. One is, that it becomes much easier to
account for the many observed continuous symmetries such as rotational and Lorentz invariance. Even more
important is the fact that a strictly continuous time coordinate implies that the hamiltonian is
unbounded, so that realistic models may be easier to achieve. But making information dissipate is not
easy in continuum theories. It may well be that discrete degrees of freedom must be added. This would be
no real problem. Discrete degrees of freedom often manifest themselves as fermions in the quantum
formalism. It is also conceivable that the continuum theories at the basis of our considerations will
have to include string- and $D$-brane degrees of freedom, and it would be beautiful if we could make more
than casual contact with the mathematics of string- and $M$-theories.

In Section~2, we expand on the definition of physical states as being equivalence classes of
deterministic states, first illustrated for the discrete case, but it has a sensible continuum limit, so
that a continuous time parameter can be employed. It is shown how dissipation of information leads to a
reduction in the number of quantum levels, but in terms of these reduced states, unitarity is restored.

In Section~3, we treat one of the continuum versions of a model with information loss, and show how they
lead to discrete quantization even if the original degrees of freedom form a continuous multi- (or
infinite-) dimensional space.

In Section~4, we show how to couple different degrees of freedom gravitationally. Gravity theory
naturally exhibits information loss when black holes are considered, and thus we argue that incorporating
the gravitational force will actually help us to understand quantum mechanics.

An axample of a non-quantum theory that could be considered for use as an input is a liquid obeying the
Navier Stokes equation, and developing turbulent behaviour. Viscosity induces information loss. The
ultraviolet structure of Navier Stokes fluids however does not quite meet the requirements of our
theories. These matters are discussed in Section~5.

The reader will criticize our arguments on the basis of the well-known Einstein-Rosen-Podolsky
paradox\ref9. We elucidate our viewpoints on this matter in Section~6. Here also we discuss an other
fundamental quantum feature of our world that may appear to be irreconcilable with a non-quantum or
pre-quantum interpretation: the `quantum computer'. Indeed we formulate a conjecture concerning the
practical limits of a quantum computer.
 
Will the Copenhagen interpretation survive the 21st century? This is discussed in Section~7. Here,
we also define the notions of {\it beables\/} and {\it changeables}. 

Dropping the requirement that information is preserved at the deterministic level, settles the problem
how to treat quantum mechanical black holes. We explain how to handle them in our theory, and what now to
think of the `holographic principle', in Sect.~8.

Conclusions are formulated in Sect.~9. In Appendix~A, we discuss the massless neutrino model, and explain
why massless neutrinos may be called quantum-deterministic objects.

\newsect{2. Quantum States} Consider a discrete system that can be in any one of the states $e_1$, $e_2$,
$e_3$ or $e_4$. We shall call these states {\it primordial\/} states. Let there be an evolution law such
that after every time step, $$e_1\ra e_2\ ,\quad e_2\ra e_1\ ,\quad e_3\ra e_3\ ,\quad e_4\ra
e_1\,.\eqno(2.1)$$ This evolution is entirely deterministic, but it will still be useful to introduce the
Hilbert space spanned by all four states, in order to handle the evolution statistically. Now, in this
space, the one-time-step evolution operator would be $$U=\pmatrix{0&1&0&1\cr 1&0&0&0\cr 0&0&1&0\cr
0&0&0&0}\,,\eqno(2.2)$$ and this would not be a unitary operator. Of course, the reason why the operator
is not unitary is that the evolution rule (2.1) is not time reversible. After a short lapse of time, only
states $e_1$, $e_2$ and $e_3$ can be reached. In this simple example, it is clear that one should simply
erase state $e_4$, and treat the upper $3\times3$ part of Eq. (2.2) as the unitary evolution matrix.
Thus, the quantum system corresponding to the evolution law (2.1) is three-dimensional, not
four-dimensional.

\midinsert\cl{\epsffile{gthjump.ps}} \scrunch{\cl{Fig.~1. The transitions of Eq.~(2.1).}} \endinsert

In more complicated non-time-reversible evolving systems, however, the `genuine' quantum states and the
false ones (the ones that cannot be reached from the far past) are actually quite difficult to
distinguish, so it is more fruitful to talk of {\it equivalence classes}. Two states are called
equivalent if, after some finite time interval, they evolve into the same state. The system described
above has three equivalence classes, $$E_1=\{e_1\}\ ,\quad E_2=\{e_2,\ e_4\}\ ,\quad
E_3=\{e_3\}\,.\eqno(2.3)$$ and the evolution operator in terms of the states $E_1,\ E_2,\ E_3$ is
$$U=\pmatrix{0&1&0\cr 1&0&0\cr 0&0&1}\,.\eqno(2.4)$$ One may consider constructing a hamiltonian operator
$H$ such that $U=e^{-iH}$. Our model universe (2.1) would be in an eigenstate of this hamiltonian. Since
the phases of the states $|e_i\ket$ are arbitrary anyway, our universe can be assumed to be either in the
state $|E_3\ket$, or in $$|\j^+\ket={1\over\sqrt2}(|E_1\ket+|E_2\ket)\,.\eqno(2.5)$$ Global time is not a
directly observable quantity (time translations can be regarded as being gauge transformations), which is
why only $|\j^+\ket$ is a physical state, together with $|E_3\ket$. So, the physical Hilbert space is
only two dimensional:$\{|\j^+\ket,\ |E_3\ket\}$.

In a system with discrete time coordinates, the hamiltonian has a periodic energy spectrum, and it is
impossible to identify any of the energy states as the true `ground state', or vacuum. This was found to
be a major obstacle impeding the construction of physically more interesting models. On the other hand,
discretization of the states is imperative for any statistical analysis. The above model now shows that,
if information is allowed to dissipate, we have to treat the equivalence classes of states as the basis
of a quantum Hilbert space, and we observe that these equivalence classes can form a smaller set than the
complete set of primordial states that one starts off with.

In the Heisenberg picture, the dimensionality of a limit cycle does not change if we replace the time
variable by one with smaller time steps, or even a continuous time. Working with a continuous time
variable has the advantage that the associated operator, the hamiltonian, is unambiguous in that case.
The hamiltonian will play a very important role in what follows.

\newsect{3. A continuum model with information loss.}

In this section, it will be shown that even in theories containing many continuous degrees of freedom,
the equivalence classes will tend to form discrete, `quantum' sets, much like the situation in the real
world, only if one allows information to dissipate. The simplest model goes as follows.

If there is a single limit cycle,  we have one periodic degree of freedom $q(t)\in [\,0,\,1\,)$, evolving
according to $$\dot q(t)=v\,,\eqno(3.1)$$ so that the period is $T=1/v$. In the Schr\"odinger picture,
the dimensionality of Hilbert space is infinite, but if this model represents an entire universe, then
only the state $E=0$ is physically acceptable. Also the fact that time is not a gauge-invariant notion
implies that only the single state $\bra q|\j\ket=1$ is physical. Therefore, in the Heisenberg picture,
the dimensionality of Hilbert space is just one.

Now imagine two such degrees of freedom:$$q_1,\,q_2\in[\,0,\,1\,)\,;\qquad\dot q_1(t)=v_1\,,\qquad\dot
q_2(t)=v_2\,.\eqno(3.2)$$ First, take $v_1$ and $v_2$ to be constants. The Schr\"odinger Hilbert space is
spanned by the states $|q_1,\,q_2\ket$, and our formal hamiltonian is $$H=v_1p_1+v_2p_2\,;\qquad
p_j=-i\pa/\pa q_j\,.\eqno(3.3)$$ In this case, even the zero-energy states span an infinite Hilbert
space, so, in the Heisenberg picture, there is an infinity of possible states.
  
Information loss is now introduced by adding a tiny perturbation that turns the flow equations into a
non-Jacobian one: $$v_1\ra v_1^0+\e f(q_1,q_2)\,;\qquad v_2\ra v_2^0+\e g(q_1,q_2)\,.\eqno(3.5)$$ The
effect of these extra terms can vary a lot, but in the generic case, one expects the following (assuming
$\e$ to be a sufficiently tiny number):

Let the ratio $v_1^0/v_2^0$ be sufficiently close to a rational number $N_1/N_2$. Then, at specially
chosen initial conditions there may be periodic orbits, with period
$$P=v_1^0/N_1=v_2^0/N_2\,,\eqno(3.6)$$ where now $v_1^0$ and $v_2^0$ have been tuned to exactly match the
rational ratio -- possible deviations are absorbed into the perturbation terms. Nearby these stable
orbits, there are non-periodic orbits, which in general will converge into any one of the stable ones,
see Fig.~2. After a sufficiently large lapse of time, we will always be in one of the stable orbits, and
all information concerning the extent to which the initial state did depart from the stable orbit, is
washed out. Of course, this only happens if the Jacobian of the evolution, the quantity $\sum_i(\pa/\pa
q_i)\dot q_i$, departs from unity. Information loss of this sort normally does not occur in ordinary
particle physics, although of course it is commonplace in macroscopic physics, such as the flow of
liquids with viscosity (see Sect.~5).

\midinsert\cl{\epsffile{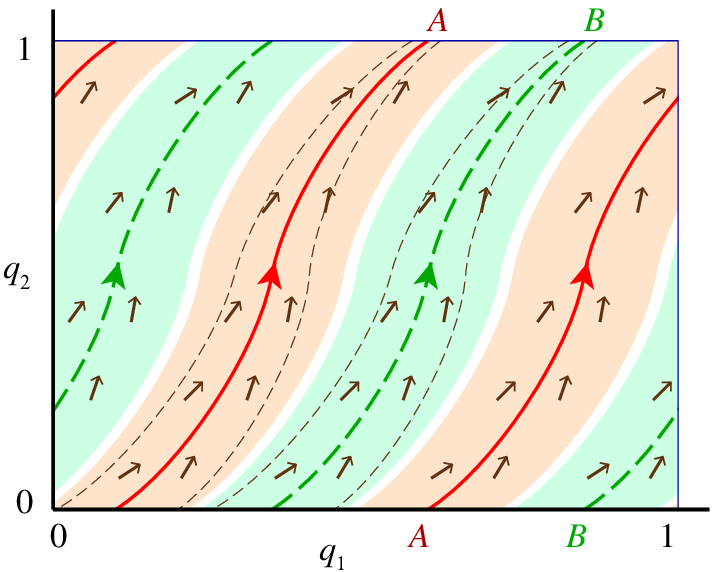}} \scrunch{\Narrower Fig.~2. Flow chart of a continuum model with two
periodic variables, $q_1$ and $q_2$. In this example, there are two stable limit cycles, $A$ and $B$,
representing the two `quantum states' of this `universe'. In between, there are two orbits that would be
stable in the time-reversed model.\Endnarrower}\endinsert

The stable orbits now represent our equivalence classes (note that, under time reversal, there are new
stable orbits in between the previous ones). Most importantly, we find that the equivalence classes will
form a discrete set, in a model of this sort, most often just a finite set, so that, in the Heisenberg
picture, our `universe' will be just in a finite number of distinct quantum states.

Generalizing this model to the case of more than two periodic degrees of freedom is straightforward. We
see that, if the flow equations are allowed to be sufficiently generic (no constraints anywhere on the
values of the Jacobians), then distinct stable limit orbits will arise. There is only one parameter that
remains continuous, which is the global time coordinate. If we insert $H|\j\ket=0$ for the entire
universe, then the global time coordinate is no longer physically meaningful, as it obtains the status of
an unobservable gauge degree of freedom.

Observe that, in the above models, what we call `quantum states', coincides with Poincar\'e limit cycles
of the universe. Just because our model universes are so small, we were able to identify these. When we
glue tiny universes together to obtain larger and hence more interesting models, we get much longer
Poincar\'e cycles, but also much more of them. Eventually, in practice, sooner or later, one has to
abandon the hope of describing complete Poincar\'e cycles, and replace them by the more practical
definitions of equivalence classes. At that point, when one combines mututally weakly interacting
universes, the effective quantum states are just multiplied into product Hilbert spaces.

\newsect{4. Gravity.}

Models describing only a small number of distinct quantum states, such as all of the above, do not very
clearly show the most salient difficulty encountered when one attempts to construct realistic models.
This is the fact that our universe is known to be {\it thermodynamic stable\/}. A system in thermodynamic
equilibrium is governed by Boltzmann factors $e^{-\b E_i}$, where $\b$ is the inverse temperature.
Stability is guaranteed only if the hamiltonian has a ground state. In the models above, only the zero
eigenvalue of the hamiltonian plays a role, so we have to be more careful in our use of the notion of a
hamiltonian. A thermodynamic treatment applies only to a hamiltonian describing some small subsystem of
the universe. Apparently, one first must address the notion of {\it locality\/} before being able to
formulate the exact meaning of thermodynamic stability. Our definition of locality will be that the
hamiltonian of the universe can be written as $$H=\int\dd^3{\bf x}\,\HH({\bf x})\,,\eqno(4.1)$$ where
$\HH({\bf x})$ is a hamiltonian density, obeying $$[\HH({\bf x}),\,\HH({\bf x'})]=0\quad\hbox{if}\quad
|{\bf x}-{\bf x'}|>\e \,,\eqno(4.2)$$ for some $\e>0$. {\it Positivity\/} then means that $\HH({\bf x})$
is bounded from below: $$\HH({\bf x})>-\d\,,\eqno(4.3)$$ for some number $\d$. The value of $\d$ may
diverge as $\e$ is sent to zero, but we do not really need to send $\e$ all the way to zero; probably a
small finite value will be good enough for us.

At first sight, it seems to be easy to realise (4.1)--(4.3) in a deterministic cellular automaton
model\ref7. In such a model, $\HH({\bf x})$ depends on only a finite number of states, and there is some
freedom in defining $\HH$, since the time variable is discrete. If the evolution law of the automaton is
local, one naturally expects that the hamiltonian will be local as well, but unfortunately, the
hamiltonian generated by a cellular automaton is not so simple. The point is that the hamiltonian is the
logarithm of the evolution operator, and it can only be written as an infinite perturbation series in
terms of the interactions. In calculating the outcome, one discovers divergences that contradict
(4.1)--(4.3), and this is the only reason why our automaton models failed to serve as realistic models
for a quantum field theory.\ref7

We propose to circumvent these problems, by first returning to continuous time variables. As we see from
the model of the previous section, it, in principle, relies on strictly continuous time, so there is a
unique hamiltonian. Finiteness of the number of quantum states is then guaranteed by the mechanism
explained, which is that information loss reduces the dimensionality of Hilbert space to be finite. The
model one starts off with, may have continuous degrees of freedom, such as a classical field theory, but
it must have information loss. We now propose that one takes a continuous, classical field theory with
general coordinate invariance. Such models differ in various essential ways from the more naive cellular
automaton models studied previously.

First, one observes that the time variable has now become a {\it local\/} gauge degree of freedom. The
velocity of time evolution in various regions of the universe may differ, just as in the model of
Sect.~3, and this difference is controlled by the gravitational field. The ratio of the speed of
evolution at $\bf x$ and ${\bf x'}$ is $\sqrt{g_{00}({\bf x})/g_{00}({\bf x'})}$, and since
$\sqrt{g_{00}}$ plays the role of a gravitational potential, this relative speed depends on the
gravitational flux from $\bf x$ to $\bf x'$. Indeed, the coupling introduced in Sect.~3 may be regarded
as a `gravitational' coupling.

Secondly, also the notion of locality is made more complicated as also the coordinates $\bf x$ have
become gauge degrees of freedom. This makes our study of the positivity constraint of the hamiltonian
density much more difficult than before.

A third complication is the exact definition of what an hamiltonian actually is. We should distinguish
the matter part of the hamiltonian from the gravitational part. It is the matter part which we want to be
positive. The gravitational part, controlling the value of the gravitational fields, must be regarded
separately, since in total they add up to zero: the hamiltonian density generates local time
translations, which however are pure gauge transformations, under which the wave function does not
change, hence $$\HH_{\rm matter}({\bf x})+\HH_{\rm grav}({\bf x})=0\,.\eqno(4.6)$$

But ``matter" should also include gravitons. The correct way to introduce the hamiltonian here is first
to define Cauchy surfaces of equal time, and then to define the operator $H$ that generates time
evolution, that is, a mapping from one Cauchy surface to the next. This requires an external clock to be
defined; we take as our clock the measurements made by observers far from the region studied. It is
important, however, that this clock is part of the universe studied, and not external to it. It is the
hamiltonian with this more subtle definition that has to be split into hamilton density functions
$\HH({\bf x})$.

All of these aspects make gravity so much different from ordinary cellular automaton models that we have
good hopes that the naive difficulties encountered with the cellular automata can now be resolved.

The most important distinction between gravitational and non-gravitational models is that, in
gravitational models, information loss naturally occurs, since black holes may be formed. Indeed, it will
be hard to avoid the development of coordinate singularities, but quite generally, one expects such
singularities to be hidden behind horizons. So we have black holes. One may wonder whether black holes in
our deterministic gravity models can emit any Hawking radiation\ref{10}, since the latter is considered
to be a typical quantum effect. The answer is that we do expect Hawking radiation, and the argument for
this is that the usual derivation of this effect is still valid. One then may ask how it can be that a
black hole can loose weight, since in classical theories black holes can only grow. The answer here is,
presumably, as follows.

In gravity, the hamiltonian not only generates the evolution through the hamilton equations, but it is
also the source of the gravitational field. In our deterministic model, the gravitational field already
exists, whereas the logarithm of the evolution operator, at first sight, may have little to do with
curvature. In writing the hamiltonian as an integral over hamiltonian densities, however, these two
notions of energy get intertwined, and we end up with only one notion of energy. When a black hole looses
energy, it is primarily because it absorbs negative amounts of ``curvature energy". Clearly then, our
primordial model must allow for the presence of negative amounts of energy. Actually, this is obviously
true for the quantum mechanical energy, because, after diagonalization, the total Hamitonian has a zero
eigenvalue. Prior to diagonalization of the total $H$, the hamilton density $\HH({\bf x})$ must have
negative eigenstates. We now see that, since the black hole must loose weight, the primordial model must
also have local fluctuations with negative ``curvature energy". Black holes absorb negative amounts of
energy, allowing positive energy to escape to infinity.

It is due to the postulated thermodynamics stability that the fluctuations surviving at spatial infinity
may only have positive energy. Since the total energy balances out, the black hole will therefore receive
only net amounts of negative energy falling in. Hence it looses weight and decays.

\newsect{5. Viscosity.}

To obtain some insight in continuum models with information loss, it is tempting to consider an example
from macroscopic physics. Consider the Navier-Stokes equations for a fluid with viscosity\ref{11}. For
simplicity, we take a pure, incompressible fluid  with density $\r$ equal to one, and viscosity $\eta$.
As is well known, such fluids may develop turbulence, and turbulence occurs when the Reynolds number, $$
R=\r u\ell/\eta\,,\eqno(5.1)$$ where $u$ represents the velocities involved and $\ell$ the typical sizes,
becomes larger than a certain critical value, $R_{\rm cr}$. This is a dimensionless number, ranging
between a few factors of 10 to something of the order of $10^3$.

Turbulence could be a nice example of the kind of chaotic behaviour to which one could apply our quantum
mechanical philosophy. We see from the expression (5.1) for Reynold's number, that only if the viscosity
$\eta$ is sufficiently small compared to the dimensions of the system, instabilities arise that cause
turbulence. Viscosity, for incompressible fluids, can be expressed in terms of $\rm cm^2/sec$, so the
distance scale at which turbulence can take place can be arbitrarily small, provided that the time scale
decreases accordingly. This is why turbulence can cascade down to very tiny dimensions, until finally the
molecular scale is reached, at which point the fluid equations no longer apply. Because of this
divergence into the infinitesimally small scales, a viscous fluid cannot be treated with our Hilbert
space methods.

In a relativistic classical field theory, the situation is likely to be very different. First of all, it
is very difficult to introduce viscosity in a relativistically invariant way, since first derivatives in
time must be linked to second derivatives in space. But, assuming that in sufficiently complicated
systems, viscous yet Lorentz invariant terms can be introduced, one notices that there must be another
distinction as well: if turbulence cascades down to smaller dimensions, it cannot be that the square of
the distance scale divided by the time scale stays constant, because the limit of the ratio of the
distance scale itself and the time scale is limited by the speed of light. Therefore, one may imagine
that there is a lower limit to vortex size, and hence a natural smallest distance limit. It is necessary
to have a smallest scale limit so as to have a workable cut-off leading to an effective quantization.
Unfortunately, realistic relativistic classical field theories with viscosity were not (yet?) found,
which is why perhaps information loss {\it via\/} black holes must be called upon.

\newsect{6. The EPR paradox. A falsifiable prediction}

The most serious objection usually raised against ideas of the kind discussed in this paper, is that
deterministic theories underlying quantum mechanics appear to imply Bell's inequalities for stochastic
phenomena\ref{12}, whereas it is well-known that many of these inequalities are violated  in quantum
mechanics. Clearly, we have to address these objections.

Bell's inequalities follow if one assumes deterministic equations of motion to be responsible for the
behaviour of quantum mechanical particles at large scales. If one assumes that the $x$-component of an
electron's spin exists, having some (unknown) value even while the $z$-component is measured, then the
usual clashes our found. In our theory, however, the wave function has exactly the meaning and
interpretation as in usual quantum mechanics; it describes the probability that something will or will
not happen, given all other information of the system available to us. ``Reality", as we perceive it,
does not refer to the question whether an electron went through one slit or another. It is our belief
that the true degrees of freedom are not describing electrons or any other particles at all, but
microscopic variables at scales comparable to the Planck scale. Their fluctuations are chaotic, and no
deterministic equation exists at all that describes the effects of these fluctuations at large scales.
Thus, the behaviour of the things we call electrons and photons is essentially entirely unpredictable. It
so happens, however, that some regularities occur within all these stochastic osscillations, and the {\it
only\/} way to describe these regularities is by making use of Hilbert space techniques.

When we measure the spin of a photon, or the detection rate of particles by a counter, our measuring
device is as much a chaotic object as the phenomena measured, and only at macroscopic scales can we
detect statistical regularities that can in no other way be linked to microscopic behaviour than by
assuming Schr\"odinger's equation. The idea that there might exist a deterministic law of physics
underlying all of this essentially amounts to nothing more than the suggestion that there exists a
`primordial basis', a preferred basis of states in Hilbert space with the property that any operator that
happens to be diagonal in this basis, will continue to be diagonal during the evolution of the system.
{\it None\/} of the operators describing present-day atomic and subatomic physics will be completely
diagonal in this basis. This enables us to accept {\it both\/} quantum mechanics with its usual
interpretation {\it and} to assume that there is a deterministic physical theory lying underneath it.

Apparently, we are forced to deny the existence of electrons, and other microscopic objects, even if they
appear to be obvious explanations of observed phenomena. Only macroscopic oscillations, such as the
movements of planets and people, are undeniable realities (that is, approximately diagonal in the
primordial basis), and it must be possible to recognise these `realities' in terms of the microscopic,
deterministic variables. This leads once again to a very serious objection, which is the following.

Quantum mechanics, as we know it, leads to many more phenomena that are at odds with classical
determininistic descriptions. An example of this is the so-called quantum computer\ref{13}. Using quantum
mechanics, a device can be built that can handle information in a way no classical machine will ever be
able to reproduce, such as the determination of the prime factors of very large numbers in an amount of
time not much more than what is needed to do multiplications and other basic arithmetic with these large
numbers. If our theory is right, it should be possible to mimick such a device using a classical theory.
This gives us a falsifiable prediction: \Narrower{\it It will never be possible to construct a `quantum
computer' that can factor a large number faster, and within a smaller region of space, than a classical
machine would do, if the latter could be built out of parts at least as large and as slow as the
Planckian dimensions.} \Endnarrower\ni A somewhat stronger version of this prediction, based on the
entropy formula for a black hole, would be: \Narrower{\it ``The classical machine may be thought of as
being built of parts each of which occupy an area of at least one Planck length squared."}\Endnarrower\ni
If this would be true, it would not be the total volume but the total area thet needs to be compared. We
are less confident, however, of this latter version of our prediction, which is based on the holographic
principle. The reason to doubt it is that the holographic principle follows from quantum mechanical
arguments, hence refers to the number of equivalence classes, not the number of actual possible states,
see Sect.~8. Therefore, a classical computer that is able to erase information, may have to use sites of
Planckian dimension in a volume, not just on an area.

Quantum computers are known to suffer from problems such as `decoherence'. Often, it is claimed that
decoherence is nothing but an annoying technical problem. In our view, it will be one of the essential
obstacles that will forever stand in the way of constructing super powerful quantum computers, and unless
mathematicians find deterministic algorithms that are much faster than the existing ones, factorization
of numbers with millions of digits will not be possible ever.

\newsect{7. Beables and changeables. Will the Copenhagen interpretation survive the 21st century?}

In a way, our present approach does not really attack the Copenhagen interpretation. We attach to the
wave function $|\j\ket$ exactly the same interpretation as the one taught at our universities. However,
the Copenhagen interpretation also carries a certain amount of agnosticism: {\it We will never be able to
determine what actually happened\/} during a physical experiment, and it is asserted that a deterministic
theory is impossible. It is this agnosticism that we disagree with. There is a single `reality', and
physicists may be able to identify some of it. Of course, our physical universe is far too complex ever
to be able to pinpoint in detail the actual sequence of events at tiny distance scales, but this
situation is in no way different from our inability to follow individual atoms and molecules in a
classical theory for gases and liquids. In a classical theory, we know that the atoms and molecules are
there, we know their dynamics, but we are unable to trace individual entities, nor are we even interested
in doing so; what we do want is to unravel the laws.

Thus, we add the following to the Copenhagen interpretation. In our theory, the \hbox{\it operators\/}
used for describing a system, will be divided in two types. If the representation of our Hilbert space is
chosen to be such that the equivalence classes of the primordial states are chosen to form its basis
elements, then we have {\it beables}, which, if expressed in this basis, multiply a state by a real or
complex number referring to properties of the equivalence class our state is in, and {\it changeables},
which may replace a state by a different state, in a different equivalence class, possibly multiplying it
also by a complex, state dependent, amplitude. Beables are operators which, in the Heisenberg
representation, all commute with one another at all times. Changeables of course do not commute, in
general. Operators that act non trivially on the different states within one equivalence class, are
physically not meaningful, but could be used for mathematical purposes. We propose to refer to these as
{\it ghost operators}. Conventional quantum mechanics results from the remarkable feature that, in
describing systems of atomic sizes, we have become unable to distinguish the beables from the
changeables. All operators known in the Standard Model of elementary particles are changeables. Beables
may refer to features at the Planck scale, or to features at macroscopic scales, but in general they are
not suitable for describing single particles at the atomic scale. Only if we manage to demonstrate that,
under several restrictive conditions, diagonalizing a beable at the macroscopic scale corresponds to
diagonalizing a changeable at the atomic scale, can we do a quantum experiment.

This, the author believes, does not contradict any of the usual findings concerning hidden variable
theories and Bell's inequalities. These findings were based on the assertion that a theory describing the
fluctuations at atomic scales should `explain' these fluctuations in terms of laws {\it at the atomic
scale} that go beyond ordinary quantum mechanics. In contrast, we now require such laws to exist {\it
only\/} at the Planck scale. It will be the physicists' task in the next century to identify the beables
that can be used at the Planck scale. They can clearly not include operators resmbling the ones we are
used to at present, such as spin, positions or momenta. At the Planck scale, the introduction of the wave
function will be nothing other than a mathematical trick enabling us to handle the equations
statistically. Due to the powerful mathematics of linear algebras, this trick will allow us to perform
renormalization goup transformations towards the much lower energy scales and much larger distance scales
of atomic physics. As a result, conventional quantum mechanics is the {\it only\/} way to describe the
correlations at atomic scales.

Our theory does profoundly disagree with the so-called `many world interpretation'. The unobserved
outcomes of experiments are not realized in `parallel universes' or anything of the sort. Every
experiment has a single outcome that is true, and all other outcomes are not realized anywhere. The wave
function only means something when it it used as a tool helping us to make statistical predictions. At
atomic and molecular scales, it is the only tool we have; there will never be a better way to make
predictions, but this does not mean that there will not be a better underlying theory.

We have no idea whether the Copenhagen interpretation, and in particular its agnostic elements, will
survive the new century or not. This depends on human ingenuity which is impossible to predict. String
theories and related approaches at present do not address at all the possibility of deterministic
underlying equations. This does not mean that they would be wrong. It is quite conceivable that string
theory is the only way to analyse our world to such detail that the underlying dynamical equations can be
identified. Our paper is a plea not to give up common sense while doing so.

\newsect{8. Black holes and holography.}

Dropping the requirement that information is preserved at the deterministic level, also settles an other
vexing problem: the treatment of quantum mechanical black holes. The problem encountered in studying the
theory for black holes is as follows.

Any sensible theory of matter and gravitation inevitably predicts that, given a sufficiently large amount
of matter, gravitational collapse may occur and a black hole may form\fn{To see this, it is sufficient to
study the Chandrasekhar limit.}. Consider now a large black hole. Its properties at moderately small
scales, can be deduced unambiguously from invariance under general coordinate transformations. An
elementary outcome of these considerations is that black holes emit particles of all kinds, in the form
of thermal radiation. This result allows us to estimate the total number of possible quantum states of a
black hole, and one finds that this number is essentially governed by the total area $A$ of the black
hole horizon\ref{10}.

On the other hand, one can try to make a model of the black hole horizon, in order to attempt to describe
these quantum states, in a statistical manner, in terms of local degrees of freedom residing at this
horizon\ref{14}. If quantum field theory is applied -- {\it any\/} quantum field theory for particles in
the background metric defined by the black hole -- one finds the number of quantum states near the
horizon to be strictly infinite. The difficulty is, that the quantum states of a field theory reside in a
volume, not on a surface, and furthermore, the number of quantum states in a field theory is unlimited
because of the freedom to perform unlimited Lorentz transformations at the extreme vicinity of a horizon.

It could be observed that what was needed is a `holographic principle'\ref4. This principle states that
the number of quantum states of the quantum field theory describing our world should not at all be as
large as in conventional, non-gravitational systems; this number should, in fact, be bounded by an
expression involving the total area of the boundary. This situation resembles what one gets if a
holographic picture is taken of a scene in three spacelike dimensions, using a two-dimensional
photographic plate. We give the photographic plate a resolution limited by one pixel per Planck length
squared, approximately. This causes a slight blurring of our three-dimensional view, but, since it is
Planckian dimensions that are involved, such a blurring is unobservable in ordinary physics. However, it
appeared to be extremely difficult to construct a theory with `holography' from first principles.

At this point, string theory and $D$-brane theory appear to come to the rescue\ref{15}: beautiful studies
provide for descriptions of black holes where, indeed, the quantum states are identified, counted, and
their number is found to depend on the horizon area in a way that was expected from Hawking radiation.
There appears to be just a small price to be paid. These theories do not tell us exactly how to handle
the space-time transformations that relate the behaviour at a horizon to theories in the nearby volume.
There are conjectures that describe the nature of these relations, but the physical implications of these
conjectures are difficult to grasp.

String theory now asserts that a theory in $3+1$ dimensions must be equivalent to a conformal theory in
lower dimensions\ref3; this has to be the case if black holes are to be adequately described by these
theories. However, it does raise all sorts of questions. In the real physical world, the number of
space-time dimensions can be determined `experimentally', and the outcome of such experiments should be
either $3+1$ or $2+1$ or something else, but not two or more conflicting answers, except in the
uninteresting case where inhabitants of this world cannot do their experiments because there are no
usable inter particle interactions, or because the interactions in their world are severely non-local
(Note that we are not referring to Kaluza-Klein compactification at this point, which of course would be
an acceptable way to link theories with different dimensionalities). How can we get around these
problems?

The theory in this paper gives a way out that is quite acceptable from a physical point of view. In our
theory, quantum states are not the primary degrees of freedom. The primary degrees of freedom are
deterministic states. Since, at a local level, information in these states is not preserved, the states
combine into equivalence classes. By construction then, the information that distinguishes the different
equivalence classes is absolutely preserved. Quantum states are equivalence classes, but in order to
identify equivalence classes, the evolution of a system must be followed for a certain length of time,
and this turns the definition of an equivalence class into a non-local one.

Black holes are nothing but extreme situations where information gets lost. Their equivalence classes
comprise large sets of states that do look quite different for a local, `infalling' observer, and this is
why a black hole contains much fewer quantum states than the world seen by someone going in. But, as
black holes are now truly large scale, composite objects, they cease to present elementary problems; they
will take care of themselves in a natural manner; what remains to be done is the determination of the
microscopic laws.

Just as all other structures in our theory, black holes will have to be described in terms of equivalence
classes of states. States that have a different past, but identical future, will be joined in a single
equivalence class. By construction, the evolution of these equivalence classes will be unitary, so the
emerging description of black hole evolution will be as in standard quantum mechanics, but the exact
formulation of the Rindler space transformation can only be given after the set of fundamental,
primordial states for the vacuum fluctuations have been identified. The so-called `holographic principle'
will then turn out to be a feature of the effective quantum mechanical description of black holes, but is
no longer needed for the description of the fundamental (deterministic) degrees of freedom of the world.
What the holographic principle tells us is, that the number of equivalence classes of the deterministic
theory will grow proportionally to the area of a black hole.

The fact that the number of equivalence classes depends only on the surface of the boundary may seem to
be something quite natural, At the boundary, information can pour in and out. If we would keep the
boundary fixed (including te vacuum fluctations there), the finite system at the inside may eventually
loose {\it all\/} of its information and turn into a single Poincar\'e cycle (or into one of a small set
of options). At closer inspection, however, this argument turns out to be insufficient. More
investigation is needed for the mechanism that reduces the number of classes to an expression depending
only on the area.

\newsect{9. Conclusions and remarks.}

In spite of the failure of macroscopic hidden variable theories, it may still be possible that the
quantum mechanical nature of the phenomenological laws of nature at the atomic scale can be attributed to
an underlying law that is deterministic at the Planck scale but with chaotic effects at all larger
scales. In this picture, what is presently perceived as a wave function must be regarded as a
mathematical device for computing probabilities for correlations in the chaos. This wave function does
retain its usual Copenhagen interpretation, but identifying quantum states at the Planck scale will be
impeded by the phenomenon of information loss at that scale. Due to information loss, Planck scale
degrees of freedom must be combined into equivalence classes, and it is these classes that will form a
special basis for Hilbert space, which we refer to as the `primordial basis'.

These considerations are of special importance for the description of black holes. The general coordinate
transformation that underlies the definition of Rindler space, maps local degrees of freedom into local
degrees of freedom. However, the fact that all information that disappeared into black holes must be
considered as being lost,  implies that the Rindler space transformation does not transform equivalence
classes into equivalence classes, and therefore, this transformation is not a transformation of quantum
states into quantum states.

Let us stress again that information loss in black holes only occurs at the classical level. Since,
according to our philosophy, quantum states are identified with equivalence classes, quantum information
is preserved, by construction. In our theory, however, we reestablish the fundamental nature of the {\it
classical\/} states, and deprive the quantum states of their fundamental  status of primary degrees of
freedom. This way, the black hole information paradox may be resolved. The well-known black hole entropy
formula tell us that the number of equivalence states for a black hole will grow as the exponent of the
area in Planck units.

It is of interest to observe that, in constructing models with a deterministic interpretation for quantum
states, the restriction to $1+1$ dimensions is usually quite helpful. This is a reason to suspect that a
deterministic interpretation of string theory is possible. In Appendix A, a construction is shown. Here,
we succeeded in producing a model in $3+1$ dimensions, but its  ultraviolet cut-off is fairly artificial.
In $1+1$ dimensions, the cut-off is straightforward.

\newsect{Appendix A. Massless neutrinos are deterministic.}

There is one system, actually realised to some extent in the real world, for which a {\it primordial
basis\/} can be constructed. A primordial basis is a complete set of basis elements of Hilbert space that
is such that any operator that happens to be diagonal now, will continue to be diagonal in the future.
Only if there is no information loss, the evolution of these elements is determined by local equations.
The model constructed in this Appendix is first constructed in such a way that no information loss
appears to occur, but it also appears to be not quite local. Then we restore locality (at the cost of a
violation of Lorentz invariance) by introducing information loss (whether Lorentz invariance has to be
broken in the real world, remains to be seen).

Consider massless, non-interacting chiral fermions in four space-time dimensions. We can think of
neutrinos, although of course real neutrinos deviate slightly from the ideal model described here.

First, take the first-quantized theory. The hamiltonian for a Dirac particle is $$H={\vec \a}\cdot{\vec
p}+\b m\,,\qquad\{\a_i,\,\a_j\}=2\d_{ij}\,,\quad \{\a_i,\,\b\}=0\,,\quad \b^2=1\,.\eqno(A.1)$$ Taking
$m=0$, we can limit ourselves to the subspace projected out by the operator $\half(1+\g_5)$, at which
point the Dirac matrices become two-dimensional. The Dirac equation then reads $$H={\vec \s}\cdot{\vec
p}\,,\eqno(A.2)$$ where $\s_{1,\,2,\,3}$ are the Pauli matrices. We now consider the basis in which the
following `primordial observables' are diagonal: $$\big\{\,\hat p\ ,\quad \hat p\cdot {\vec \s}\ ,\quad
\hat p\cdot\vec  x\,\big\}\ ,\eqno(A.3)$$ where $\hat p$ stands for $\pm\vec p/|p|$, with the sign such
that $\hat p_x>0$. We do {\it not\/} directly specify the sign of $\vec p$.

Writing $p_j=-i\part{ }{x_j}$, one readily checks that these three operators commute, and that they
continue to do so at all times. Indeed, the first two are constants of the motion, whereas the last one
evolves into $$\hat p\cdot\vec x(t)=\hat p\cdot\vec x(0)+\hat p\cdot\vec\s\,t\,.\eqno(A.4)$$

The fact that these operators are complete is also easy to verify: in momentum space, $\hat p$ determines
the orientation; let us take this to be the $z$ direction. Then, in momentum space, the absolute value of
$p$, as well as its sign, are identified with its $z$-component, and it is governed by the operator
$i\pa/\pa p_z=x_z=\hat p\cdot\vec x$. The spin is defined in the $z$-direction by $\hat p\cdot\vec\s$.

Mathematically, these equations appear to describe a {\it plane}, or a flat membrane, moving in
orthogonal direction with the speed of light. Given the orientation (without its sign) $\hat p$, the
coordinate $\hat p\cdot\vec x$ describes its distance from the origin, and the variable $\hat
p\cdot\vec\s$ specifies in which of the two possible orthogonal directions the membrane is moving. Note
that, indeed, this operator flips sign under $180^\circ$ rotations, as it is required for a spin $\half$
representation. This, one could argue, is what a neutrino really is: a flat membrane
moving in the orthogonal direction with the speed of light. But we'll return to that later: the theory
can be further improved.

We do note, of course, that in the description of a single neutrino, the Hamiltonian is not bounded from
below, as one would require. In this very special model, there is a remedy to this, and it is precisely
Dirac's second quantization procedure. We consider a space with an infinite number of these membranes,
running in all of the infinitely many possible directions $\hat p\cdot\vec\s$. In order to get the
situation under control, we introduce a temporary cut-off: in each of the infinitely many possible
directions $\hat p$, we assume that the membranes sit in a discrete lattice of possible positions. The
lattice length $a$ may be as small as we please. Furthermore, consider a box with length $L$, being as
large as we please. The first-quantized neutrino then has a finite number of energy levels, between
$-\pi/a$ and $+\pi/a$. The state we call `vacuum state', has all negative energy levels filled and all
positive energy levels empty. All excited states now have positive energy. Since the Dirac particles do
not interact, their numbers are exactly conserved, and the collection of all observables (A.3) for all
Dirac particles still correspond to mutually commuting operators.

In this very special model we thus succeed in producing a complete set of primordial observables, {\it
i.e.\/}, operators that commute with one another at all times, whereas the hamiltonian is bounded from
below. We consider this to be an existence proof, but it would be more satisfying if we could have
produced a less trivial model. Unfortunately, our representation of neutrinos as infinite, strictly flat
membranes, appears to be impossible to generalise so as to introduce mass terms and/or interactions.
Also, the flat membranes appear to be irreconcilable with space-time curvature in a gravity theory.
Quite likely, one has to introduce information loss. Suppose we may drop Lorentz invariance for the
deterministic underlying theory\fn{At a later stage, this could lead to a tiny, in principle detectable,
violation of Lorentz invariance for the quantum system.}. We then may add as physical variables also
transverse coordinates $\tl x$, orthogonal to $\hat p$. Particles are now described in terms of all three
space coordinates $\vec x$, and a direction operator $\hat p$ (reabsorbing $\hat p\cdot\vec\s$ to
indicate its sign). In the direction of $\hat p$, the propagation is rigid. But in the orthogonal
direction, the propagation is haphazard, such that information concerning the initial value of $\tl x$ is
lost. All states with the same $\hat p$ and $\hat p\cdot \vec x$, but with different $\tl x$, will have
to be put in the same equivalence class. Thus, it is the equivalence classes that form flat membranes,
while the deterministic theory may be strictly local.

\newsect{References}

\item{1.}M.B. Green, J.H. Schwarz and E. Witten, {\it Superstring Theory}",  Cambridge 
     Univ. Press.
\item{2.} P.K.~Townsend, in {\it Frontiers in Quantum Physics}, Kuala Lumpur 1997, S.C.~Lim et al, Eds., Springer
1998, p. 15.
\item{3.}E.~Witten, {\it Anti de Sitter Space and holography}, hep-th/9802150; 
 J.~Maldacena, {\it The large $N$ Limit of superconformal field theories and supergravity}, hep-th/9711020;
 T.~Banks et al, {\it Schwarzschild Black Holes from Matrix Theory}, hep-th/9709091;
 K.~Sken\-deris, {\it Black holes and branes in string theory}, SPIN-1998/17, hep-th/9901050.
 \item{4.} G. 't Hooft, {\it Dimensional reduction in quantum gravity.} In {\it 
Salamfestschrift: a collection of talks}, World Scientific Series in 20th Century 
Physics, vol. {\bf 4}, Eds. A. Ali, J. Ellis and S. Randjbar-Daemi (World Scientific, 
1993), THU-93/26, gr-qc/9310026;  {\it Black holes and the dimensionality of space-time}, in 
Proceedings of the Symposium ``The Oskar Klein Centenary'', 19-21 Sept. 1994,
Stockholm, Sweden. Ed. U. Lindstr\"om, World Scientific 1995,  p. 122;
L.~Susskind,  {\it J.~Math.~Phys. \bf 36} (1995) 6377, hep-th/9409089.
\item{5.}A.~Staruszkiewicz, {\it Acta Phys. Polon. \bf 24} (1963) 734; 
     S.~Giddings, J.~Abbott and K.~Kuchar, {\it Gen.~Rel. and Grav. \bf  16}  (1984) 
     751; S.~Deser, R.~Jackiw and G.~'t Hooft, {\it Ann.~Phys. \bf 152} (1984) 220;
J.R.~Gott, and M.~Alpert, {\it Gen.~Rel.~Grav. \bf 16} (1984) 243;  
 J.R.~Gott, {\it Phys.~Rev.~Lett.} {\bf 66} (1991) 1126;
 S.~Deser, R.~Jackiw and G.~'t Hooft, {\it Phys.~Rev.~Lett.} {\bf 68} (1992) 267;
S.M.~Carroll, E.~Farhi and A.H.~Guth, {\it Phys.~Rev.~Lett.} {\bf 68} (1992) 263;
 G.~'t Hooft,  {\it Class.~Quantum Grav. 
 \bf 9} (1992) 1335.
\item{6.} A.~Achucarro and P.K.~Townsend, {\it Phys.~Lett. \bf B180} (1986) 89; 
E.~Witten, {\it Nucl.~Phys.} {\bf B311} (1988) 46;
S.~Carlip, {\it Nucl.~Phys.} {\bf B324} (1989) 106, and in: "Physics,  Geometry  and 
Topology", NATO ASI series B, Physics, Vol. {\bf 238}, H.C.~ Lee  ed.,  Plenum 
1990, p.~541;
J.E.~Nelson and T.~Regge, {\it
Quantisation of 2+1 gravity for genus 2}, Torino prepr. DFTT 54/93,
gr-qc/9311029;
G.~'t Hooft, {\it Class.~Quantum Grav.} {\bf 10} (1993) 1023, {\it
ibid.} {\bf 10} (1993) S79; {\it {\it Nucl.~Phys.}} {\bf B30} (Proc.~Suppl.)
(1993) 200;
S.~Carlip, {\it Six
ways to quantize (2+1)-dimensional gravity}, Davis Preprint UCD-93-15,
gr-qc/9305020;
G.~Grignani, {\it 2+1-dimensional gravity as a gauge
theory of the Poincar\'e group}, Scuola Normale Superiore, Perugia,
Thesis 1992-1993; 
 G.~'t Hooft, {\it {\it Commun.~Math.~Phys.}} {\bf 117}
(1988) 685;   {\it Nucl.~Phys.} {\bf B342} (1990) 471, 
Class.~Quantum Grav.~{\bf 13}  1023;
S.~Deser and R.~Jackiw, Comm.~Math.~Phys. {\bf 118} (1988) 495.   
\item{7.} G.~'t Hooft, {\it J.~Stat.~Phys. \bf 53} (1988) 323; {\it Nucl.~Phys. \bf B342} (1990) 471; 
G.~'t Hooft,  K.~Isler and S.~Kalitzin, {\it Nucl.~Phys.} {\bf B 386} (1992) 495.
\item{8.} G.~'t Hooft, {\it Quantummechanical behaviour in a deterministic model},
{\it Foundations of Physics letters \bf 10} no. 4, April 1997, quant-ph/9612018. 
\item{9.}A.~Einstein, B.~Podolsky and N.~Rosen, {\it Phys.~Rev.} {\bf  47} (1935) 777.  
\item{10.} S.W.~Hawking, {\it Commun.~Math.~Phys.} {\bf 43} (1975) 199;
J.B.~Hartle and S.W.~Hawking, Phys.Rev. {\bf D13} (1976) 2188.                         
\item{11.} See e.g. L.D.~Landau and E.M.~Lifshitz, Course of Theoretical Physics, Vol 6, {\it  Fluid Mechanics},
Pergamon Press, Oxford 1959.
\item{12.} J.S.~Bell, Physica {\bf 1} (1964) 195.
\item{13.} A.~Eckert and R.~Jozsa, {\it Rev.~Mod.~Physics \bf 68} (1996) 733.
\item{14.} G.~'t Hooft, {\it Nucl.~Phys.} {\bf B256} (1985) 727.
\item{15.} J.~Polchinski, {\it Phys.~Rev.~Lett. \bf 75} (1995) 4724, hep-th/9510017.
 
\bye